\documentclass[aps,pre,twocolumn,showpacs,superscriptaddress,groupedaddress]{revtex4}
\usepackage{graphicx}
\usepackage{bm}        
\usepackage{amssymb}   
\begin{document}

\title{Chiral Structure of F-actin Bundle Formed by Multivalent Counterions }

\author{Sarah Mohammadinejad}
\email{sarah@iasbs.ac.ir}
\affiliation{Department of Physics, Institute for Advanced Studies in Basic Sciences (IASBS), Zanjan 45137-66731, Iran}
\affiliation{Department of Biological Sciences, Institute for Advanced Studies in Basic Sciences (IASBS), Zanjan 45137-66731, Iran}

\author{Ramin Golestanian}
\email{ramin.golestanian@physics.ox.ac.uk}
\affiliation{Rudolf Peierls Centre for Theoretical Physics, University of Oxford, Oxford OX1 3NP, UK.}

\author{Hossein Fazli}
\email{fazli@iasbs.ac.ir}
\affiliation{Department of Physics, Institute for Advanced Studies in Basic Sciences (IASBS), Zanjan 45137-66731, Iran}
\affiliation{Department of Biological Sciences, Institute for Advanced Studies in Basic Sciences (IASBS), Zanjan 45137-66731, Iran}

\date{\today}

\begin{abstract}
The mechanism of multivalent counterion-induced bundle formation by filamentous actin (F-actin)
is studied using a coarse-grained model and molecular dynamics simulation. Real diameter size,
helically ordered charge distribution and twist rigidity of F-actin are taken into account in our model. The attraction between parallel F-actins induced by multivalent counterions is studied in detail and it is found that the maximum attraction occurs between their closest charged domains. The model F-actins aggregate due to the like-charge attraction and form closely packed bundles. Counterions are mostly distributed in the narrowest gaps between neighboring F-actins inside the bundles and the channels between three adjacent F-actins correspond to low density of the counterions. Density of the counterions varies periodically with a wave length comparable to the separation between consecutive G-actin monomers along the actin polymers. Long-lived defects in the hexagonal order of F-actins in the bundles are observed that their number increases with increasing the bundles size. Combination of electrostatic interactions and twist rigidity has been found not to change the symmetry of F-actin helical conformation from the native $\frac{13}{6}$ symmetry. Calculation of zero-temperature energy of hexagonally ordered model F-actins with the charge of the counterions distributed as columns of charge domains representing counterion charge density waves has shown that helical symmetries commensurate with the hexagonal lattice correspond to local minima of the energy of the system. The global minimum of energy corresponds to $\frac{24}{11}$ symmetry with the columns of charge domains arranged in the narrowest gaps between the neighboring F-actins.
\end{abstract}

\maketitle

\section{Introduction}\label{Sec-Intro}
F-actin is a highly charged polyelectrolyte that plays a key role in the cytoskeleton of eukaryotic cells,
where it forms a network of bundles made by crosslinking proteins \cite{Molecular2002Albert}.
It has also been observed that multivalent counterions can cause these macromolecules to attract each other (the phenomenon known as like-charge attraction)
and form aggregates \cite{Direct2005Angelini,LikeCharge2004Angelini,Lamellar2003Wong}.

When the separation between two polyelectrolytes like F-actin is considerably larger than their diameter, they can be modeled as uniformly charged lines, cylinders, or bead-spring chains, for the calculation of their electrostatic interaction potential
\cite{Counterion1997Ha,Wigner1999Shklovskii,Counterion1997Gronbech,Electrostatic2002YLevin,Equilibrium2005Henle,Chirality2007Grason,
Parametrization1994Podgornik,Electrostatic1999Kornyshev,Interaction1968Oosawa,An1994Ray,Symmetries2010Landy,Azimuthal2003Harreis,
Bundle1999Steven,Effect2004Lee,Attraction2003Deserno,Orientationally2009Mohammadinejad,Salt2009Fazli,Aggregation2007Fazli,Finite2007Sayar,Equilibrium2010Sayar,
Cohesive2003Rudnick}.
Despite their simplicity, such coarse-grained minimal models have been shown to suffice for obtaining the essence of the overall behavior of these polyelectrolytes in solution. Analytical and computational studies using these models have shown that the interaction potential between two similarly charged stiff polyelectrolytes in multivalent salt solution crucially depends on their separation, mutual orientation and the concentration and valence of the salt \cite{Effect2004Lee,Ion2003Butler,Structrul2005Borukhov,Raft2001Borukhov,Orientationally2009Mohammadinejad}. Many aspects of the collective behavior of stiff polyelectrolytes in multivalent salt solution, such as the formation of aggregates of various
structures and high sensitivity of these structures to the salt concentration and valence, are believed to originate from
the complexity and high anisotropy of their interaction potential \cite{Effect2004Lee,Structrul2005Borukhov,Raft2001Borukhov,Orientationally2009Mohammadinejad}. The complexity is the result of
the combination of the highly anisotropic shape of stiff polyelectrolytes and the simultaneous presence of short- and long-range interactions in the system, and is also known to lead to frustration and slow dynamics \cite{Orientational2005Fazli,Dynamics1997Muthukumar}.

The interaction between these macromolecules at close separations, however, cannot be studied using such simplistic models.
In this case, the detailed structure of the macromolecules such as their helical symmetry, charge distribution, thickness, dielectric constant, twist rigidity and so on, will need to be taken into consideration. It has been observed that like-charge attraction between helical polyelectrolytes such as F-actin and DNA induced by multivalent salt could organize them in closely packed, hexagonally ordered bundles of parallel polyelectrolytes. Native helical symmetry of these polyelectrolytes may be incommensurate with the mentioned hexagonal order\cite{Cooperativity2009Shin,Structural2010Shin,DNA1996Bloomfield,Symmetry1998Kornyshev}. In this case, the interplay between twist rigidity of the polyelectrolytes and electrostatic correlations determines their equilibrium conformation in the bundle. There have been a number of theoretical studies on interaction between helical macromolecules in the literature \cite{Theory1997Kornyshev,Electrostatic1999Kornyshev}. Experimental study of bundle formation by F-actin in the presence of multivalent salt has shown that there are major differences between F-actin bundles formed in this way and those formed non-electrostatically \cite{LikeCharge2004Angelini}.  To study like-charge attraction between such polyelectrolytes in close separations the model should contain details of the charge distribution on the polyelectrolytes and their orientational and torsional degrees of freedom. Coarse-grained models (such as the bead-spring model) are not suitable for such studies. On the other hand, computer simulation of these macromolecules in atomistic scale is too time consuming, while a large number of the internal degrees of freedom will also be redundant and have no role in the interaction between the macromolecules. Therefore, a sufficiently detailed coarse-grained model that incorporates the helical structure of the macromolecules seems to be the most appropriate starting point for such a study.

In this paper, we study the mechanism of like-charge attraction between coarse-grained model F-actins in the presence of multivalent counterions using molecular dynamics (MD) simulations. We find for two parallel F-actins that the maximum attraction occurs between monomers whose charged domains are in the closest separation. The attractive potential of mean force between F-actins reveals no detectable dependence on their helical symmetry. The like-charge attraction between F-actins organizes them in bundles of hexagonal lattice structure containing long-lived defects. The number of the defects in a bundle is found to be an increasing function of its size, namely the number of F-actins forming the bundle. In the cross section of each bundle, the low- and the high-counterion-density regions are observed that correspond to the narrowest gaps between neighboring F-actins and the centers of triangles formed by centers of three adjacent F-actins, respectively. Competition between electrostatic interactions and twist rigidity of F-actins appears to be unable to change their native helical conformation of $\frac{13}{6}$ symmetry into structures with other symmetries (the definition of F-actin helical symmetry is given below, see Sec. \ref{Sec-Model}). The helices of the F-actins are observed to form domains of uniform phase that are connected via defects analogous to those in 2D XY model. Considering the charge of counterions distributed as columns of frozen charge domains similar to the counterion charge density waves introduced in Ref. \cite{LikeCharge2004Angelini}, we calculate zero-temperature energy of the system as a function of the helical symmetry of the F-actins. We find that there are numerous local minima in the energy of the system corresponding to helical symmetries commensurate with the hexagonal order of the F-actins in the bundle. The ground state of the system corresponds to the $\frac{24}{11}$ helical symmetry of F-actins when the columns of charged regions are at the midpoints between the axes of neighboring F-actins.

The rest of the paper is organized as follows. The model and the simulation method are introduced in Sec. \ref{Sec-Model}. The study of like-charge attraction between a pair of F-actins is presented in Sec. \ref{Sec-LikeCharge}. In Sec. \ref{Sec-BundleMechanism}, the counterion density profile, as well as the positional and helical structures of the filaments in the bundle are studied. The helical symmetry of F-actins in the bundle is studied in Sec. \ref{Sec-Ideal} using the simplified smeared counterion distribution model.
Section \ref{Sec-Conclusion} concludes the paper.

\section{The model and the simulation method}\label{Sec-Model}
\begin{figure}[b]
\centering
\resizebox{0.99\columnwidth}{!}{
  \includegraphics{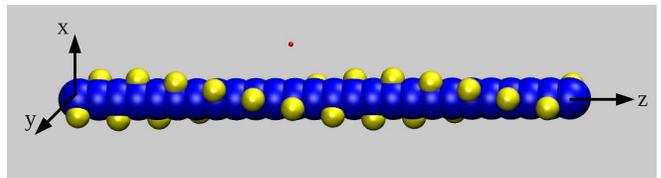}}
  \caption{The model F-actin constructed by two kinds of spheres. The large overlapping blue spheres model F-actin backbone and carry no charge. The yellow spheres carry all the net negative charge of G-actin monomers. The small red sphere (whose diameters are scaled by a factor of three to make them visible) represent the counterions. The spatial order of the yellow spheres around the backbone is determined from the real helical order of F-actin and the structure of G-actin monomer obtained from X-ray experiments \cite{Atomic1990Holm,Structure1993McLaughlin,Structural1995AlKhayat}.}
  \label{Fig-ActinModel}
\end{figure}

We use a coarse-grained model for F-actin that takes into consideration a number of key macromolecular features such as its diameter (that is relatively large, say as compared to the mobile ions), helical order of its charge distribution, persistent length, rotational and twist degrees of freedom, and its twist rigidity. We construct the model F-actin by assembling two kinds of spheres (see Fig. \ref{Fig-ActinModel}): a large neutral sphere that makes the backbone of F-actin (that represents the volume of actin) and a smaller negatively charged sphere that carry the net charge of F-actin (which has a linear charge density $\lambda_{F-actin}=\frac{-1e}{2.5\AA}$). Each G-actin monomer is thus modeled as a set of a large sphere and an attached
small charged sphere. The adjacent large spheres are bonded to each other by a harmonic potential, $\frac{U_h}{k_BT}=\frac{1}{2}k_h(r-r_0)^2$, of spring constant $k_h=500 \sigma^{-2}$, where $r_0$ is the average separation between them and $\sigma$ is the MD length scale (see below). Three-particle angle potential, $\frac{U_{bend}}{k_BT}=k_{bend}(1-\cos \theta)$, with $k_{bend}=1000$ is used to mimic F-actin bending rigidity and keep our model F-actins as rod-like polymers (since their lengths are considerably smaller than the persistence length of F-actin).
The positions of the small spheres are determined from X-ray observation-based four-sphere model of G-actin \cite{Atomic1990Holm,Structure1993McLaughlin,Structural1995AlKhayat}. These spheres play the role of subdomain-1 (sd1), which is known to carry
most of the G-actin charge. According to the four-sphere model, the position of the sd1 domain of the $n$th
monomer in the regular conformation of F-actin, can be given by
\begin{eqnarray}
\label{Eq_Factin_helix}
\nonumber \rho&=&28.33\AA\\
\phi_n&=&\phi_1+(n-1)\phi_0\\
\nonumber z_n&=&z_1+(n-1)z_0,
\end{eqnarray}
in cylindrical coordinate system. Here, $z_1$ and $\phi_1$ are the height and the azimuthal coordinates of sd1 of the first monomer and $\phi_0$ and $z_0$ are the increments of $\phi$ and $z$ coordinates from each monomer to the next one. We set $z_0=28.7 \AA$ and determine
the value of $\phi_0$ from the helical symmetry of F-actin, such that for helical symmetry of $\frac{m}{n}$, $\phi_0=-\frac{n\times 2\pi}{m}$.  For example, for helical symmetry of $\frac{13}{6}$, $\phi_0=-\frac{6\times 2\pi}{13}=-166.2^\circ$. Small spheres in the model F-actin are bonded to the large ones by a harmonic potential of spring constant $k_h$. To model twist rigidity of F-actin \cite{Torsional1996Tsuda} ($\kappa_{twist}=8\times 10^{-26} Nm^2$), the dihedral potential $\frac{U_{twist}}{k_BT}=k_{twist}\left[1-\cos(\phi- \phi_0)\right]$ is applied between two neighboring monomers with $k_{twist}=6900$, where $\phi-\phi_0$ is deviation of dihedral angle between two consecutive monomers from its native value, $\phi_0$. From the known charge density of F-actin, it can be calculated that each small sphere carries the charge of $q=-11e$ ($e$ is the elementary charge). The charge of each small sphere is considered as a point charge in its center. We explicitly add 3-valent counterions of charge $3e$ as Lennard-Jones (LJ) particles of diameter $\sigma=2.2\AA$ (MD length scale in our simulations) to neutralize the charge of F-actins. Lennard-Jones potential also introduces an energy scale, $\epsilon$, to the system. We use 3-valent (not 2-valent) counterions to strengthen the electrostatic correlations in the system and probe the role of these correlations easily. One should note that because of the coarse-graining method in our model there is no essential difference between 2- and 3-valent counterions.

Considering that in typical actin bundling experiments \cite{LikeCharge2004Angelini} the salt concentration is 36-108 mM, electrostatic interactions are strongly screened and act only at short ranges (the screening length is $\lambda_{\rm Debye}\simeq 10\AA$).
Like-charge attraction between polyelectrolytes is known to originate from correlations and fluctuations of multivalent counterions in their vicinity.
To account for the effect of salt and avoid the time consuming calculations of the long-ranged Coulomb interactions, we use a screened Debye-H\"{u}ckel potential and explicitly add counterions to the system. The Debye-H\"{u}ckel potential for the electrostatic interaction between the charged particles $i$ and $j$ is given as $\frac{U_{DH}}{k_BT}=l_B\frac{Z_iZ_j}{r_{ij}}e^{-\kappa r_{ij}}$, where $l_B\simeq 7\AA$ is the Bjerrum length (in water at room temperature), $Z_i$($Z_j$) is the valence of charged particle $i$($j$) and $r_{ij}$ is the separation between charges $i$ and $j$. We use $\kappa^{-1}=10\sigma$. One should note that effects such as electrostriction which may have some contribution \cite{Solvent2008Noe,How2002Hribar,Peptide2009Tulip} are not considered here. We denote the numbers of F-actins, 3-valent counterions and monomers of each F-actin by $N_p$, $N_c$, and $N_m$, respectively. Our MD simulations are performed with the MD simulation
\begin{figure}[h]
\centering
\resizebox{0.9\columnwidth}{!}{
  \includegraphics{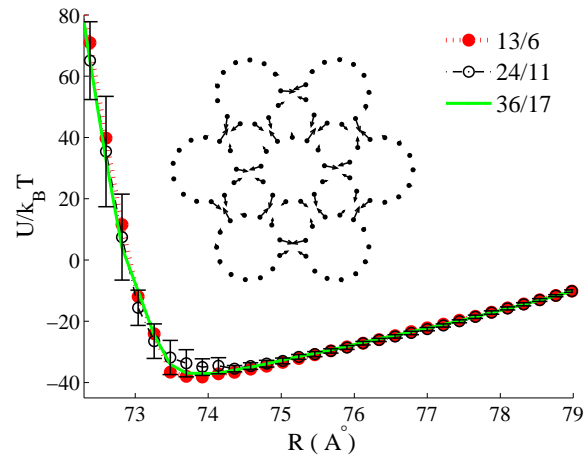}}
  \caption{The potential of mean force per monomer between a pair of parallel F-actins of the native helical symmetry and two other symmetries, $\frac{24}{11}$ and $\frac{36}{17}$, in the presence of 3-valent counterions as a function of their separation, $R$. Each data point is obtained from averaging over $10^5$ integration time steps after equilibration of the system. Error bars which are shown only on data points of $\frac{24}{11}$ symmetry are obtained from averaging over 6 pairs (central F-actin with 6 others surrounding it). As it can be seen, the values of the potential of mean force for the three helical symmetries are the same within the error bars. Inset: Top view of the configuration of 7 F-actins and average forces acting on their monomers for a given value of $R$.}
  \label{Fig-PotMeanForce}
\end{figure}
package ESPResSo \cite{Espresso2006Limbach}. Periodic boundary condition (PBC) is applied and a Langevin thermostat with friction coefficient $\Gamma=1.0 \sqrt{\frac{\epsilon m}{\sigma^2}}$ is used to keep the temperature fixed at $k_BT=\epsilon$ ($m$ is the mass of the mobile ions). The simulation box is rectangular and its length in $z$ direction is equal to the length of F-actins to minimize the end effects. The lengths of the box in the other two directions are the same and determined by considering the cut-off length of the screened electrostatic interaction and the number of F-actins in the system. Truncated and shifted LJ potential,
\begin{equation}
 U_{LJ}(r) = \left\lbrace
  \begin{array}{l l}
    4\epsilon\
    \left[\left(\frac{\sigma}{r-r_o}\right)^{12}-\left(\frac{\sigma}{r-r_o}\right)^{6}+\frac{1}{4}\right] & \text{if $r<r_o+r_{c}$},\\
    0 & \text{if $r \geq r_o+r_{c}$},
  \end{array}
\right. \label{Eq-LJ}
\end{equation}
is used to model excluded volume interactions. $r_{c}=\sqrt[6]{2}$ and $r_o$ are the cut-off and offset of this potential\cite{Espresso2006Limbach}, respectively. In the cases of backbone and sd1 spheres of model F-actins, the corresponding diameters of the spheres, namely $\sigma_b=28.33\AA\times 2=56.66\AA$ and $\sigma_{sd1}=16.81\AA\times 2=33.62\AA$, are used to set the value of the variable $r_0$ in the LJ potential. In the case of counterion-counterion excluded volume interactions, $r_o=0$. LJ time scale in our simulations is $\tau_0=\sqrt{\frac{m\sigma^2}{\epsilon}}$ and the integration time step is $\tau=0.02\tau_0$.

\section{Like-charge attraction between two parallel F-actins}\label{Sec-LikeCharge}

To calculate the potential of mean force between a pair of parallel model F-actins in the presence of 3-valent counterions, we first fix $N_p=7$ of them parallel to each other on a hexagonal lattice of separation $R$ as shown in the inset of Fig. \ref{Fig-PotMeanForce}. Keeping all of their degrees of freedom blocked, we leave the counterions to equilibrate. We then release the rotation and twist degrees of freedom ($\phi_1$ and $\phi_0$ in Eq. \ref{Eq_Factin_helix}) of all F-actins, such that they can explore their equilibrium conformations. After equilibration of the system, we calculate the time average of the interaction forces between the central F-actin and 6 others surrounding it. From integration of the mean force (averaged over time and over 6 pairs of F-actins) with respect to the separation $R$, we calculate the potential of mean force for a pair of F-actins. The potential of mean force per monomer versus the separation is shown in Fig. \ref{Fig-PotMeanForce}.
We calculate the potential of mean force for three different values of $\phi_0$ corresponding to helical symmetries $\frac{13}{6}$, $\frac{24}{11}$, and $\frac{36}{17}$. Our results show that the interaction potential as a function of the separation does not depend on F-actins helical symmetry and no detectable difference for above-mentioned symmetries is observed (see Fig. \ref{Fig-PotMeanForce}). We also observe that the difference between initial phases, $\phi_1$, of the interacting F-actins tends to vanish in equilibrium.

\begin{figure}[h]
\centering
\resizebox{0.9\columnwidth}{!}{
  \includegraphics{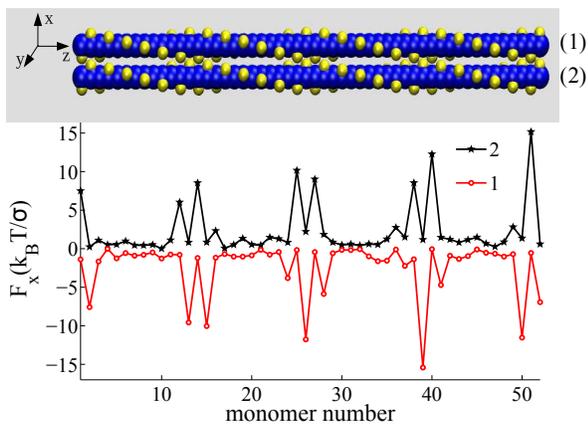}}
  \caption{The time average of the forces acting on monomers of two parallel model F-actins in the presence of 3-valent counterions. Equilibrium configuration of the F-actins is shown in the upper panel. The strongest attractions are between the monomers whose charged domains are closest to each other. The values of the variable $\phi_1$ for the two F-actins are the same.}
  \label{Fig-ForceAlong}
\end{figure}

To understand the mechanism of interaction between a pair of neighboring model F-actins it is instructive to look at the forces acting on each of their monomers. The average forces acting on monomers of two interacting F-actins are shown in Fig. \ref{Fig-ForceAlong}, which shows that the mutual attraction is strongest between the few monomers in each helical period whose charged domains are closest to each other. Interactions between the other monomers are found to be considerably smaller than that of the overlapping charged domains. The few overlapping domains produce a strong attractive potential for the counterions and gather them in the narrow space between these domains.

\section{Bundle formation by F-actins and its mechanism}\label{Sec-BundleMechanism}

To study the bundle formation of the filaments, we first fix $N_p$ parallel F-actins of length $N_m=26$ on a hexagonal lattice of spacing $R=80 A^\circ$,
larger than the value corresponding to the minimum of the potential of mean force (see Fig. \ref{Fig-PotMeanForce}). We then leave the
counterions to equilibrate, and subsequently release all the degrees of freedom of the F-actins, except that we constrain the two ends of the filament to move on the two boundary surfaces of the simulation box. We have also repeated the same procedure with non-hexagonal initial configurations of F-actins. We observe that regardless of the initial configuration of the F-actins they attract each other and form a bundle in which their axes form a hexagonal lattice. Moreover, the initial phases of the helices (variable $\phi_1$ in Eq. \ref{Eq_Factin_helix}) are observed to approach each other in the course of the aggregation dynamics.

\begin{figure}[t]
\centering
\resizebox{0.8\columnwidth}{!}{
  \includegraphics{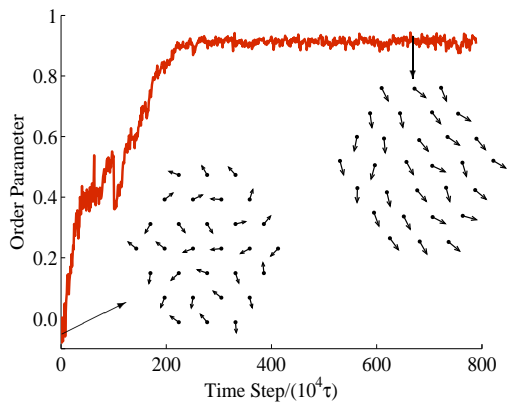}}
  \caption{The order parameter, $X$, defined in Eq. \ref{Eq-X}, as a function of time obtained from simulation of $N_p=30$ model F-actins. Through directional ordering of the phase vectors the order parameter changes from zero to a saturation near unity. Typical snapshots of the phase vectors are shown for randomly oriented and orientationally ordered configurations.}
  \label{Fig-OrderPar}
\end{figure}

To visualize this behavior we represent the initial phase, $\phi_1$, of each F-actin by a vector directed from F-actin axis to sd1 of its first monomer and probe the directional ordering of these vectors in the $x-y$ plane in our simulations. We perform these simulations for different system sizes, $N_p=7-52$. In the case of $N_p=7$, the system easily finds its low-energy state in which parallel F-actins are ordered hexagonally and their phase vectors are mostly aligned in the same direction. For large $N_p$, however, usually there are some defects in the hexagonal order of F-actins in the bundle, with their number increasing when the number of F-actins in \begin{figure}[h]
\centering
\resizebox{0.9\columnwidth}{!}{
  \includegraphics{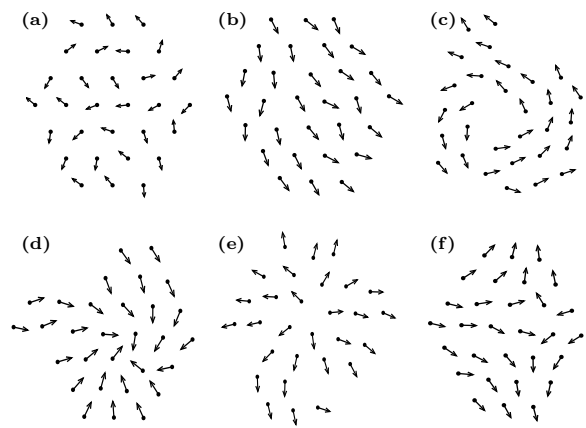}}
  \caption{a) Initial configuration of $N_p=30$ parallel F-actins (top view) with non-ordered phase vectors. b-f) Five typical configurations of F-actins after running the system for a long time ($\simeq 5\times 10^6$ integration time steps). b) Both positional and directional orders are approximately complete. c-f) Different kinds of defects in F-actins positional hexagonal order and directional order of their phase vectors.  }
  \label{Fig-XYSourceSink}
\end{figure}
the system increases. These defects are very long-lived and for the systems larger than $N_p\simeq 10$ it is not possible to observe their disappearance and to find defect-less hexagonal order in the system in our simulations. Moreover, defects in the long-range directional order of the phase vectors (similar to those found in 2D XY model) are also observed in the system (see Fig. \ref{Fig-XYSourceSink}). The positional and the orientational defects are strongly correlated. Namely, when there are positional defects in the system, certainly orientational defects are also present. Orientational defects are also observed in systems without any positional defects. Low-energy state of the system corresponds to a configuration in which F-actins form a hexagonally ordered bundle with no defect and with their phase vectors all aligned in the same direction.
We define an order parameter, $X$, as
\begin{equation}
X=\Sigma_{<ij>} \cos (\phi_1^{i}-\phi_1^{j}) ,
\label{Eq-X}
\end{equation}
in which the summation runs over nearest neighbor pairs to probe directional ordering of the phase vectors. A typical behavior of the order parameter, $ X $, as a function of time is shown in Fig. \ref{Fig-OrderPar}. The figure shows that the order parameter starts from the initial value of zero corresponding to complete lack of order in the phase vectors orientation, and saturates to a value that represents the degree of alignment between the phase vectors.

Simulations of F-actins with the value of $\phi_0$ set to the native $\frac{13}{6}$ symmetry show no appreciable change in the F-actins helical symmetry upon bundle formation. Although local deviations of dihedral angles between adjacent monomers along F-actins are observed, the average value of the dihedral angles agrees well with the native value. We performed these simulations with two other helical symmetries of the model F-actins, namely $\frac{36}{17}$ and $\frac{24}{11}$, and observed no noticeable changes in the results. Statistics and averages of deviations of dihedral angles between adjacent monomers from their native values for three different helical symmetries are the same within the error bars. We also performed simulations with model F-actins with native helical symmetry of $\frac{13}{6}$ but using initial conformations (after equilibration of the counterions) that had symmetries of $\frac{36}{17}$ and $\frac{24}{11}$ (two separate simulations). We found that in both cases, after releasing the F-actins degrees of freedom their symmetry returned back (in a short time) to the native $\frac{13}{6}$. These observations suggest that electrostatic correlations are not capable of keeping changing the helical symmetry of F-actins in bundle.

\begin{figure}[t]
\centering
\resizebox{0.8\columnwidth}{!}{
  \includegraphics{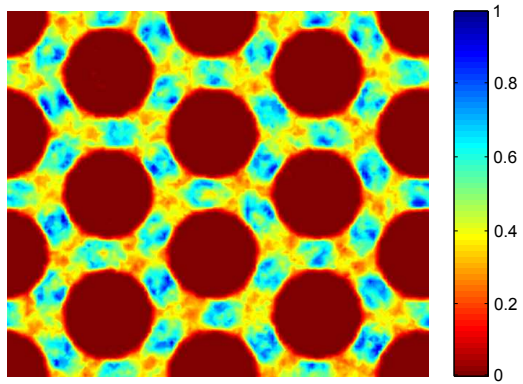}}
  \caption{Top view of counterion density profile inside a hexagonally ordered bundle of $N_p=30$ model F-actins (a part of the system is shown here). The counterion density has its maximum value in the middle of the separation between the axes of neighboring F-actins. The centers of triangles formed by three neighboring F-actins have the lowest value of the counterion density.}
  \label{Fig-XYDistCount}
\end{figure}

To study the counterion distribution and the force field acting on G-actin monomers in a bundle of hexagonally ordered parallel F-actins, we construct a bundle of $N_p$ fixed F-actins with their axes on a hexagonal lattice and leave the counterions to equilibrate. We choose the lattice constant as the value that corresponds to the minimum of the potential of mean force shown in Fig. \ref{Fig-PotMeanForce}. We then release the twist and rotational degrees of freedom of F-actins and allow them to equilibrate. After equilibration of the system we probe the counterion distribution and the forces acting on the monomers of F-actins and any deviation of the twist order of F-actins from its native symmetry. A typical counterion density profile chosen from a part of the system containing $N_p=30$ model F-actins is shown in Fig. \ref{Fig-XYDistCount}, in which the distribution profile is obtained using the superposition of $10^3$ snapshots of the system chosen from $10^5$ integration time steps of the simulation.

\begin{figure}[t]
\centering
\resizebox{0.7\columnwidth}{!}{
  \includegraphics{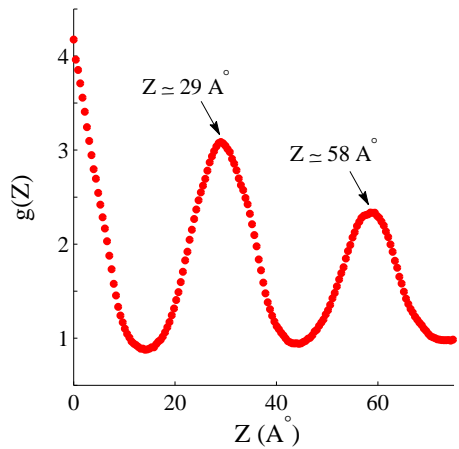}}
  \caption{Pair correlation function of the counterions along $z$ axis inside a hexagonally ordered bundle of parallel F-actins which its cross-section in $x-y$ plane is shown in Fig. \ref{Fig-XYDistCount}. As it can be seen the most repeated separation between the counterions along the F-actins is approximately equal to the separation between consecutive G-actin monomers. }
  \label{Fig-Pair_Correlation}
\end{figure}

We also calculate counterion-counterion pair correlation function in $z$ direction. This correlation function shows that the most repeated separation between between counterions in $z$ direction is the separation between sd1s of consecutive monomers along F-actins (see Fig. \ref{Fig-Pair_Correlation}). This result combined with the result shown in Fig. \ref{Fig-XYDistCount} shows that the counterions are mostly distributed in the narrowest gap between two neighboring F-actins inside the bundle as domains of separation equal to monomer-monomer separation along the F-actins. The centers of the channels formed by three adjacent F-actins correspond to a lower density of the counterions despite the picture of charge density wave presented in Ref. \cite{LikeCharge2004Angelini}.

\begin{figure}[t]
\centering
\resizebox{0.9\columnwidth}{!}{
  \includegraphics{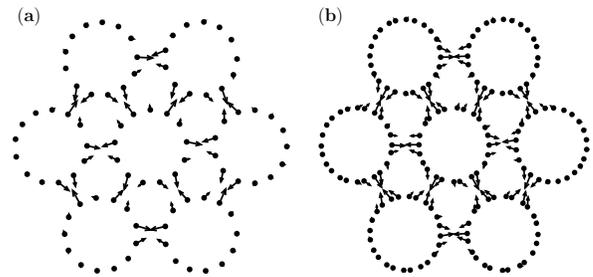}}
  \caption{Forces acting on monomers of $N_p=7$ model F-actins in the presence of 3-valent counterions (top view) averaged over $10^3$ snapshots of the system chosen from $10^5$ integration time steps for helical symmetries $\frac{13}{6}$ (a) and $\frac{24}{11}$ (b).}
  \label{Fig-ForceField}
\end{figure}

Time-average forces acting on the actin monomers in a bundle of $N_p=7$ model F-actins for two values of $\phi_0$ corresponding to $\frac{13}{6}$ and $\frac{24}{11}$ symmetries are shown in Fig. \ref{Fig-ForceField}. The main difference between helical symmetries $\frac{13}{6}$ and $\frac{24}{11}$ is related to their commensurability with the hexagonal lattice. Despite the native $\frac{13}{6}$ symmetry of F-actin, symmetries like $\frac{24}{11}$ and $\frac{36}{17}$ are commensurate with hexagonal lattice (considering its definition, symmetry $\frac{m}{n}$ is commensurate with hexagonal lattice if $m$ is divisible by $6$). As Fig. \ref{Fig-ForceField} shows, the strengths and directions of the forces that act on actin monomers are slightly more regular in the case of $\frac{24}{11}$. Apart from that, we observed no detectable difference between the time-average forces acting on the F-actins for the three different symmetries. We note that we have found some evidence suggesting that the number (or density) of positional defects for the same value of $N_p$ is higher for the native symmetry of $\frac{13}{6}$ as compared to those that are commensurate with the hexagonal order of F-actins. Although this observation seems plausible, it needs to be probed more systematically using larger systems.

\section{Hexagonally ordered F-actin bundle with columns of charge domains of opposite sign}\label{Sec-Ideal}

\begin{figure}[t]
\centering
\resizebox{0.8\columnwidth}{!}{
  \includegraphics{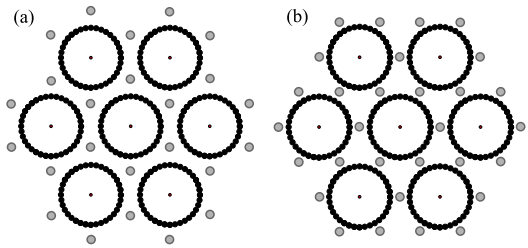}}
  \caption{Top view of two possible arrangements of the columns of charge domains inside the F-actins bundle introduced in the text. Counterion charge domains and sd1s of G-actin monomers along the F-actins are shown by gray and black circles, respectively.}
  \label{Fig-SimCodeModel}
\end{figure}

Here we consider a bundle of hexagonally ordered parallel F-actins with the neutralizing charge of the counterions distributed as equally charged domains inside the bundle. The charge domains form one dimensional lattices of spacing $2\times z_0=57.4\AA$ along F-actins and are regularly distributed inside the bundle. Considering symmetries of the system, two different arrangements of the columns of charged regions are possible. Top view of these two arrangements are shown in Fig. \ref{Fig-SimCodeModel}. The amount of the charge of these domains, which are considered as point charges in centers of gray spheres (see Fig. \ref{Fig-SimCodeModel}), are determined in a way that the total charge of the system is zero. For both of these arrangements we calculate total energy of the system composed of electrostatic interactions (Debye-H\"{u}ckel potential with $\lambda_{Debye}=22\AA$) and twist energy of F-actins, for different values of regular deviation of F-actins helical order from the native $\frac{13}{6}$ symmetry. In Fig. \ref{Fig-SimCodeEnergy}, energy of the system versus the value of the deviation from the native symmetry for the separation between neighboring F-actins, $R=60\AA$, is shown. As it can be seen, there are some local minima in
  the energy of the system corresponding to symmetries of F-actin which are commensurate with hexagonal lattice. The
global minimum corresponds to $\frac{24}{11}$ symmetry of arrangement (b) in Fig. \ref{Fig-SimCodeModel}. Also, as it can be seen in this figure, energy of the system is considerably lower and local minima are considerably deeper in the case of arrangement (b) relative to arrangement (a). A similar study for arrangement of multivalent counterions inside a bundle of polyelectrolytes modeled as uniformly charged lines has been performed \cite{Cohesive2003Rudnick}. Despite the model studied in Ref. \cite{Cohesive2003Rudnick}, helical order of F-actin charge distribution enters the subject of commensurability into the problem in our model. We also looked at the effect of the slide of the neighboring one-dimensional lattices of charge domains with respect to one another and found that such slides increase the system energy and the low energy state of the system corresponds to the configuration with no slide of these columns relative to one another. Considering these results in addition to the counterion distribution obtained from the bulk simulations at room temperature (Fig. \ref{Fig-XYDistCount}), it seems that the counterions have a high tendency to the narrowest gap between two F-actins in the bundle and the picture shown in Fig. \ref{Fig-XYDistCount} is the dominant scenario relative to the other picture in which counterions are mostly distributed inside the channels formed by three adjacent F-actins.
\section{Conclusions and discussion}\label{Sec-Conclusion}

In conclusion, using a coarse-grained model and MD simulations, the mechanism of like-charge attraction between F-actins and bundle formation in the presence of multivalent counterions have been studied. Important macromolecular features of F-actin such as its diameter size, persistence length, helically ordered charge distribution, and twist rigidity have been considered in our model. It has been shown that parallel F-actins attract each other in the presence of 3-valent counterions and the maximum attraction occurs between monomers that their charged domains are in the closest separation. Due to the like-charge attraction, F-actins aggregate and form closely packed bundles of parallel F-actins. It has been found that the phase difference between helical structure of neighboring F-actins in a bundle tends to vanish. Defects in the long-range orientational order of the phase vectors with interesting similarities to those of 2D XY model have also been observed. In addition, long-lived positional defects in the hexagonal order of F-actins in the bundles have been observed that their number increases with increasing the bundle size. Existence of similar defects have also been reported in experimental studies of similar systems \cite{Direct2005Angelini,LikeCharge2004Angelini}.
\begin{figure}[t]
\centering
\resizebox{0.8\columnwidth}{!}{
  \includegraphics{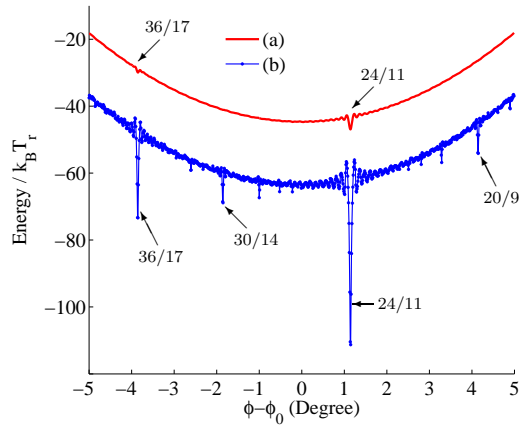}}
  \caption{Energy per monomer of the central F-actin in arrangements (a) and (b) of Fig. \ref{Fig-SimCodeModel} versus the deviation of F-actins helical symmetry from the native $\frac{13}{6}$ symmetry, $\phi-\phi_0$. Minima in the energy curve correspond to symmetries of F-actin which are commensurate with hexagonal order. The global minimum corresponds to symmetry $\frac{24}{11}$. F-actin in this symmetry is $1.2^\circ$ undertwisted relative to its native symmetry. $T_r$ refers to room temperature.}
  \label{Fig-SimCodeEnergy}
\end{figure}

Counterions inside the bundles have been observed that are mostly distributed in the narrowest gap between neighboring F-actins and the channels formed by three adjacent F-actins in the bundles contain lower density of the counterions. The density of counterions in the narrowest gap between neighboring F-actins in the bundles varies periodically along F-actins with a wave length approximately equal to the separation between consecutive G-actin monomers along the actin polymer.

Our results obtained from simulations presented in Sec. \ref{Sec-BundleMechanism} show that in the bundle formation process, deviation of dihedral angle between consecutive monomers along F-actins from its native value happens. It seems that by these local deviations the system decreases the incommensurability effects of F-actins native structure with the hexagonal order. Such local deviations from a regular symmetry in F-actin helical structure has also been reported in experimental studies \cite{Factin1982Egelman}. Changing of F-actin helical symmetry from $\frac{13}{6}$ to other symmetries has not been observed in our simulations. Positional defects of life time longer than our simulation times have been observed in the hexagonal order of F-actins in the bundles. The number of these defects has been found that increases with increasing the bundle size. Also, it has been observed that no considerable coarsening of these defects happens in our simulations because of very slow dynamics of their evolution inside the bundle. Although additional studies of these defects in larger systems are needed, existence of such defects could be a reason for finite sided bundles of F-actin in multivalent salt solution.

Study of a bundle of hexagonally ordered parallel F-actins with the charge of counterions distributed as columns of charge domains inside the bundle similar to the CDW picture of Ref. \cite{LikeCharge2004Angelini} showed that the configuration of minimum energy corresponds to $\frac{24}{11}$ helical symmetry of F-actins with columns of charged regions in the middle of the separation between axes of neighboring F-actins. In this arrangement, sd1s of F-actins and the charge domains of opposite sign are closer to each other in comparison with the other possible arrangement in which the columns of charged domains are located on the axes of the channels formed by three adjacent F-actins. This calculation shows that if we arrange the counterions according to the scenario of the charge density wave presented in Ref. \cite{LikeCharge2004Angelini}, the minimum energy of the system in our model corresponds to $\frac{24}{11}$ symmetry, not $\frac{36}{17}$.
Considering F-actin torsional modulus, overtwisting it from its native helical symmetry to for example $\frac{36}{17}$ symmetry ($3.8^\circ$ over-twist per monomer) costs an energy of $\simeq 16k_BT$ per monomer. The other symmetry, $\frac{24}{11}$, which corresponds to the global minimum of energy in Fig. \ref{Fig-SimCodeEnergy}, differs only $1.2^\circ$ per monomer from F-actin native structure. Energy cost of such a deviation (undertwist) is in the range of thermal fluctuations strength and under-twisting of F-actin to this symmetry seems to be possible. In our simulations however, we found that the system prefers the F-actins to have local deviations from their native symmetries instead of regularly changing of the helical symmetry.

Some important features of real F-actin macromolecules have not been considered in our model.
Despite our assumption, it is known that G-actin monomers charge is distributed as discrete point charges on the surface of its domains. We assumed however in our model all the net charge of G-actin as a point charge in the center of its sd1. One should note that other models containing more details of F-actin may be obtained from simulations of F-actin in the atomistic scale \cite{Structural2011Splettstoesser,Coarse2006Chu}. The other important feature of real F-actin that has not been considered in our model is the low value of its dielectric constant, $\simeq 2$, relative to that of water, $\simeq 80$. This difference is known that can affect distribution of counterions in the vicinity of such macromolecules \cite{Strong2008Jho}.
\begin{acknowledgements}
We deeply thank G. Wong, C. Holm, F. J\"{u}licher and F. Mohammad-Rafiee for enlightening discussions. We also gratefully acknowledge the helpful comments of M. A. Charsooghi, L. Mollazadeh and N. Nikoofard.
\end{acknowledgements}

\footnotesize{
\bibliography{sarahmohammadinejad} 

\begin{thebibliography}{47}
\expandafter\ifx\csname natexlab\endcsname\relax\def\natexlab#1{#1}\fi
\expandafter\ifx\csname bibnamefont\endcsname\relax
  \def\bibnamefont#1{#1}\fi
\expandafter\ifx\csname bibfnamefont\endcsname\relax
  \def\bibfnamefont#1{#1}\fi
\expandafter\ifx\csname citenamefont\endcsname\relax
  \def\citenamefont#1{#1}\fi
\expandafter\ifx\csname url\endcsname\relax
  \def\url#1{\texttt{#1}}\fi
\expandafter\ifx\csname urlprefix\endcsname\relax\def\urlprefix{URL }\fi
\providecommand{\bibinfo}[2]{#2}
\providecommand{\eprint}[2][]{\url{#2}}

\bibitem[{\citenamefont{Bruce et~al.}(2002)\citenamefont{Bruce, Alexander,
  Julian, Martin, Keith, and Peter}}]{Molecular2002Albert}
\bibinfo{author}{\bibfnamefont{A.}~\bibnamefont{Bruce}},
  \bibinfo{author}{\bibfnamefont{J.}~\bibnamefont{Alexander}},
  \bibinfo{author}{\bibfnamefont{L.}~\bibnamefont{Julian}},
  \bibinfo{author}{\bibfnamefont{R.}~\bibnamefont{Martin}},
  \bibinfo{author}{\bibfnamefont{R.}~\bibnamefont{Keith}}, \bibnamefont{and}
  \bibinfo{author}{\bibfnamefont{W.}~\bibnamefont{Peter}},
  \emph{\bibinfo{title}{{Molecular biology of the cell}}}
  (\bibinfo{publisher}{Garland Science}, \bibinfo{year}{2002}),
  \bibinfo{edition}{4th} ed.

\bibitem[{\citenamefont{Angelini et~al.}(2005)\citenamefont{Angelini, Liang,
  Wriggers, and Wong}}]{Direct2005Angelini}
\bibinfo{author}{\bibfnamefont{T.}~\bibnamefont{Angelini}},
  \bibinfo{author}{\bibfnamefont{H.}~\bibnamefont{Liang}},
  \bibinfo{author}{\bibfnamefont{W.}~\bibnamefont{Wriggers}}, \bibnamefont{and}
  \bibinfo{author}{\bibfnamefont{G.}~\bibnamefont{Wong}},
  \bibinfo{journal}{Eur. Phys. J. E} \textbf{\bibinfo{volume}{16}},
  \bibinfo{pages}{389} (\bibinfo{year}{2005}).

\bibitem[{\citenamefont{Angelini et~al.}(2003)\citenamefont{Angelini, Liang,
  Wriggers, and Wong}}]{LikeCharge2004Angelini}
\bibinfo{author}{\bibfnamefont{T.~E.} \bibnamefont{Angelini}},
  \bibinfo{author}{\bibfnamefont{H.}~\bibnamefont{Liang}},
  \bibinfo{author}{\bibfnamefont{W.}~\bibnamefont{Wriggers}}, \bibnamefont{and}
  \bibinfo{author}{\bibfnamefont{G.~C.~L.} \bibnamefont{Wong}},
  \bibinfo{journal}{Proc. Natl. Acad. Sci. USA} \textbf{\bibinfo{volume}{100}},
  \bibinfo{pages}{8634} (\bibinfo{year}{2003}).

\bibitem[{\citenamefont{Wong et~al.}(2003)\citenamefont{Wong, Lin, Tang, Li,
  Janmey, and Safinya}}]{Lamellar2003Wong}
\bibinfo{author}{\bibfnamefont{G.~C.~L.} \bibnamefont{Wong}},
  \bibinfo{author}{\bibfnamefont{A.}~\bibnamefont{Lin}},
  \bibinfo{author}{\bibfnamefont{J.~X.} \bibnamefont{Tang}},
  \bibinfo{author}{\bibfnamefont{Y.}~\bibnamefont{Li}},
  \bibinfo{author}{\bibfnamefont{P.~A.} \bibnamefont{Janmey}},
  \bibnamefont{and} \bibinfo{author}{\bibfnamefont{C.~R.}
  \bibnamefont{Safinya}}, \bibinfo{journal}{Phys. Rev. Lett.}
  \textbf{\bibinfo{volume}{91}}, \bibinfo{pages}{018103+}
  (\bibinfo{year}{2003}).

\bibitem[{\citenamefont{Ha and Liu}(1997)}]{Counterion1997Ha}
\bibinfo{author}{\bibfnamefont{B.~Y.} \bibnamefont{Ha}} \bibnamefont{and}
  \bibinfo{author}{\bibfnamefont{A.~J.} \bibnamefont{Liu}},
  \bibinfo{journal}{Phys. Rev. Lett.} \textbf{\bibinfo{volume}{79}},
  \bibinfo{pages}{1289} (\bibinfo{year}{1997}).

\bibitem[{\citenamefont{Shklovskii}(1999)}]{Wigner1999Shklovskii}
\bibinfo{author}{\bibfnamefont{B.~I.} \bibnamefont{Shklovskii}},
  \bibinfo{journal}{Phys. Rev. Lett.} \textbf{\bibinfo{volume}{82}},
  \bibinfo{pages}{3268} (\bibinfo{year}{1999}).

\bibitem[{\citenamefont{Gr{\o}nbech-Jensen
  et~al.}(1997)\citenamefont{Gr{\o}nbech-Jensen, Mashl, Bruinsma, and
  Gelbart}}]{Counterion1997Gronbech}
\bibinfo{author}{\bibfnamefont{N.}~\bibnamefont{Gr{\o}nbech-Jensen}},
  \bibinfo{author}{\bibfnamefont{R.~J.} \bibnamefont{Mashl}},
  \bibinfo{author}{\bibfnamefont{R.~F.} \bibnamefont{Bruinsma}},
  \bibnamefont{and} \bibinfo{author}{\bibfnamefont{W.~M.}
  \bibnamefont{Gelbart}}, \bibinfo{journal}{Phys. Rev. Lett.}
  \textbf{\bibinfo{volume}{78}}, \bibinfo{pages}{2477} (\bibinfo{year}{1997}).

\bibitem[{\citenamefont{Levin}(2002)}]{Electrostatic2002YLevin}
\bibinfo{author}{\bibfnamefont{Y.}~\bibnamefont{Levin}}, \bibinfo{journal}{Rep.
  Prog. Phys.} \textbf{\bibinfo{volume}{65}}, \bibinfo{pages}{1577}
  (\bibinfo{year}{2002}).

\bibitem[{\citenamefont{Henle and Pincus}(2005)}]{Equilibrium2005Henle}
\bibinfo{author}{\bibfnamefont{M.~L.} \bibnamefont{Henle}} \bibnamefont{and}
  \bibinfo{author}{\bibfnamefont{P.~A.} \bibnamefont{Pincus}},
  \bibinfo{journal}{Phys. Rev. E} \textbf{\bibinfo{volume}{71}},
  \bibinfo{pages}{060801} (\bibinfo{year}{2005}).

\bibitem[{\citenamefont{Grason and Bruinsma}(2007)}]{Chirality2007Grason}
\bibinfo{author}{\bibfnamefont{G.~M.} \bibnamefont{Grason}} \bibnamefont{and}
  \bibinfo{author}{\bibfnamefont{R.~F.} \bibnamefont{Bruinsma}},
  \bibinfo{journal}{Phys. Rev. Lett.} \textbf{\bibinfo{volume}{99}},
  \bibinfo{pages}{098101} (\bibinfo{year}{2007}).

\bibitem[{\citenamefont{Podgornik et~al.}(1994)\citenamefont{Podgornik, Rau,
  and Parsegian}}]{Parametrization1994Podgornik}
\bibinfo{author}{\bibfnamefont{R.}~\bibnamefont{Podgornik}},
  \bibinfo{author}{\bibfnamefont{D.~C.} \bibnamefont{Rau}}, \bibnamefont{and}
  \bibinfo{author}{\bibfnamefont{V.~A.} \bibnamefont{Parsegian}},
  \bibinfo{journal}{Biophys. J.} \textbf{\bibinfo{volume}{66}},
  \bibinfo{pages}{962} (\bibinfo{year}{1994}).

\bibitem[{\citenamefont{Kornyshev and
  Leikin}(1999)}]{Electrostatic1999Kornyshev}
\bibinfo{author}{\bibfnamefont{A.~A.} \bibnamefont{Kornyshev}}
  \bibnamefont{and} \bibinfo{author}{\bibfnamefont{S.}~\bibnamefont{Leikin}},
  \bibinfo{journal}{Phys. Rev. Lett.} \textbf{\bibinfo{volume}{82}},
  \bibinfo{pages}{4138} (\bibinfo{year}{1999}).

\bibitem[{\citenamefont{Oosawa}(1968)}]{Interaction1968Oosawa}
\bibinfo{author}{\bibfnamefont{F.}~\bibnamefont{Oosawa}},
  \bibinfo{journal}{Biopolymers} \textbf{\bibinfo{volume}{6}},
  \bibinfo{pages}{1633} (\bibinfo{year}{1968}).

\bibitem[{\citenamefont{Ray and Manning}(1994)}]{An1994Ray}
\bibinfo{author}{\bibfnamefont{J.}~\bibnamefont{Ray}} \bibnamefont{and}
  \bibinfo{author}{\bibfnamefont{G.~S.} \bibnamefont{Manning}},
  \bibinfo{journal}{Langmuir} \textbf{\bibinfo{volume}{10}},
  \bibinfo{pages}{2450} (\bibinfo{year}{1994}).

\bibitem[{\citenamefont{Landy and Rudnick}(2010)}]{Symmetries2010Landy}
\bibinfo{author}{\bibfnamefont{J.}~\bibnamefont{Landy}} \bibnamefont{and}
  \bibinfo{author}{\bibfnamefont{J.}~\bibnamefont{Rudnick}},
  \bibinfo{journal}{Phys. Rev. E} \textbf{\bibinfo{volume}{81}},
  \bibinfo{pages}{061918} (\bibinfo{year}{2010}).

\bibitem[{\citenamefont{Harreis et~al.}(2003)\citenamefont{Harreis, Likos, and
  L\"{o}wen}}]{Azimuthal2003Harreis}
\bibinfo{author}{\bibfnamefont{H.~M.} \bibnamefont{Harreis}},
  \bibinfo{author}{\bibfnamefont{C.~N.} \bibnamefont{Likos}}, \bibnamefont{and}
  \bibinfo{author}{\bibfnamefont{H.}~\bibnamefont{L\"{o}wen}},
  \bibinfo{journal}{Biophys. J.} \textbf{\bibinfo{volume}{84}},
  \bibinfo{pages}{3607} (\bibinfo{year}{2003}).

\bibitem[{\citenamefont{Stevens}(1999)}]{Bundle1999Steven}
\bibinfo{author}{\bibfnamefont{M.~J.} \bibnamefont{Stevens}},
  \bibinfo{journal}{Phys. Rev. Lett.} \textbf{\bibinfo{volume}{82}},
  \bibinfo{pages}{101} (\bibinfo{year}{1999}).

\bibitem[{\citenamefont{Lee et~al.}(2004)\citenamefont{Lee, Borukhov, Gelbart,
  Liu, and Stevens}}]{Effect2004Lee}
\bibinfo{author}{\bibfnamefont{K.~C.} \bibnamefont{Lee}},
  \bibinfo{author}{\bibfnamefont{I.}~\bibnamefont{Borukhov}},
  \bibinfo{author}{\bibfnamefont{W.~M.} \bibnamefont{Gelbart}},
  \bibinfo{author}{\bibfnamefont{A.~J.} \bibnamefont{Liu}}, \bibnamefont{and}
  \bibinfo{author}{\bibfnamefont{M.~J.} \bibnamefont{Stevens}},
  \bibinfo{journal}{Phys. Rev. Lett.} \textbf{\bibinfo{volume}{93}},
  \bibinfo{pages}{128101} (\bibinfo{year}{2004}).

\bibitem[{\citenamefont{Deserno et~al.}(2003)\citenamefont{Deserno, Arnold, and
  Holm}}]{Attraction2003Deserno}
\bibinfo{author}{\bibfnamefont{M.}~\bibnamefont{Deserno}},
  \bibinfo{author}{\bibfnamefont{A.}~\bibnamefont{Arnold}}, \bibnamefont{and}
  \bibinfo{author}{\bibfnamefont{C.}~\bibnamefont{Holm}},
  \bibinfo{journal}{Macromolecules} \textbf{\bibinfo{volume}{36}},
  \bibinfo{pages}{249} (\bibinfo{year}{2003}).

\bibitem[{\citenamefont{Mohammadinejad
  et~al.}(2009)\citenamefont{Mohammadinejad, Fazli, and
  Golestanian}}]{Orientationally2009Mohammadinejad}
\bibinfo{author}{\bibfnamefont{S.}~\bibnamefont{Mohammadinejad}},
  \bibinfo{author}{\bibfnamefont{H.}~\bibnamefont{Fazli}}, \bibnamefont{and}
  \bibinfo{author}{\bibfnamefont{R.}~\bibnamefont{Golestanian}},
  \bibinfo{journal}{Soft Matter} \textbf{\bibinfo{volume}{5}},
  \bibinfo{pages}{1522} (\bibinfo{year}{2009}).

\bibitem[{\citenamefont{Fazli et~al.}(2009)\citenamefont{Fazli, Mohammadinejad,
  and Golestanian}}]{Salt2009Fazli}
\bibinfo{author}{\bibfnamefont{H.}~\bibnamefont{Fazli}},
  \bibinfo{author}{\bibfnamefont{S.}~\bibnamefont{Mohammadinejad}},
  \bibnamefont{and}
  \bibinfo{author}{\bibfnamefont{R.}~\bibnamefont{Golestanian}},
  \bibinfo{journal}{J. Phys.: Condens. Matt.} \textbf{\bibinfo{volume}{21}},
  \bibinfo{pages}{424111+} (\bibinfo{year}{2009}).

\bibitem[{\citenamefont{Fazli and Golestanian}(2007)}]{Aggregation2007Fazli}
\bibinfo{author}{\bibfnamefont{H.}~\bibnamefont{Fazli}} \bibnamefont{and}
  \bibinfo{author}{\bibfnamefont{R.}~\bibnamefont{Golestanian}},
  \bibinfo{journal}{Phys. Rev. E} \textbf{\bibinfo{volume}{76}},
  \bibinfo{pages}{041801} (\bibinfo{year}{2007}).

\bibitem[{\citenamefont{Sayar and Holm}(2007)}]{Finite2007Sayar}
\bibinfo{author}{\bibfnamefont{M.}~\bibnamefont{Sayar}} \bibnamefont{and}
  \bibinfo{author}{\bibfnamefont{C.}~\bibnamefont{Holm}},
  \bibinfo{journal}{Europhys. Lett.} \textbf{\bibinfo{volume}{77}},
  \bibinfo{pages}{16001} (\bibinfo{year}{2007}).

\bibitem[{\citenamefont{Sayar and Holm}(2010)}]{Equilibrium2010Sayar}
\bibinfo{author}{\bibfnamefont{M.}~\bibnamefont{Sayar}} \bibnamefont{and}
  \bibinfo{author}{\bibfnamefont{C.}~\bibnamefont{Holm}},
  \bibinfo{journal}{Phys. Rev. E} \textbf{\bibinfo{volume}{82}},
  \bibinfo{pages}{031901} (\bibinfo{year}{2010}).

\bibitem[{\citenamefont{Rudnick and Jasnow}(2003)}]{Cohesive2003Rudnick}
\bibinfo{author}{\bibfnamefont{J.}~\bibnamefont{Rudnick}} \bibnamefont{and}
  \bibinfo{author}{\bibfnamefont{D.}~\bibnamefont{Jasnow}},
  \bibinfo{journal}{Phys. Rev. E} \textbf{\bibinfo{volume}{68}},
  \bibinfo{pages}{051902} (\bibinfo{year}{2003}).

\bibitem[{\citenamefont{Butler et~al.}(2003)\citenamefont{Butler, Angelini,
  Tang, and Wong}}]{Ion2003Butler}
\bibinfo{author}{\bibfnamefont{J.~C.} \bibnamefont{Butler}},
  \bibinfo{author}{\bibfnamefont{T.}~\bibnamefont{Angelini}},
  \bibinfo{author}{\bibfnamefont{J.~X.} \bibnamefont{Tang}}, \bibnamefont{and}
  \bibinfo{author}{\bibfnamefont{G.~C.~L.} \bibnamefont{Wong}},
  \bibinfo{journal}{Phys. Rev. Lett.} \textbf{\bibinfo{volume}{91}},
  \bibinfo{pages}{028301} (\bibinfo{year}{2003}).

\bibitem[{\citenamefont{Borukhov et~al.}(2005)\citenamefont{Borukhov, Bruinsma,
  Gelbart, and Liu}}]{Structrul2005Borukhov}
\bibinfo{author}{\bibfnamefont{I.}~\bibnamefont{Borukhov}},
  \bibinfo{author}{\bibfnamefont{R.~F.} \bibnamefont{Bruinsma}},
  \bibinfo{author}{\bibfnamefont{W.~M.} \bibnamefont{Gelbart}},
  \bibnamefont{and} \bibinfo{author}{\bibfnamefont{A.~J.} \bibnamefont{Liu}},
  \bibinfo{journal}{Proc. Natl. Acad. Sci. USA} \textbf{\bibinfo{volume}{102}},
  \bibinfo{pages}{3673} (\bibinfo{year}{2005}).

\bibitem[{\citenamefont{Borukhov and Bruinsma}(2001)}]{Raft2001Borukhov}
\bibinfo{author}{\bibfnamefont{I.}~\bibnamefont{Borukhov}} \bibnamefont{and}
  \bibinfo{author}{\bibfnamefont{R.~F.} \bibnamefont{Bruinsma}},
  \bibinfo{journal}{Phys. Rev. Lett.} \textbf{\bibinfo{volume}{87}},
  \bibinfo{pages}{158101} (\bibinfo{year}{2001}).

\bibitem[{\citenamefont{Fazli et~al.}(2005)\citenamefont{Fazli, Golestanian,
  and Kolahchi}}]{Orientational2005Fazli}
\bibinfo{author}{\bibfnamefont{H.}~\bibnamefont{Fazli}},
  \bibinfo{author}{\bibfnamefont{R.}~\bibnamefont{Golestanian}},
  \bibnamefont{and} \bibinfo{author}{\bibfnamefont{M.~R.}
  \bibnamefont{Kolahchi}}, \bibinfo{journal}{Phys. Rev. E}
  \textbf{\bibinfo{volume}{72}}, \bibinfo{pages}{011805}
  (\bibinfo{year}{2005}).

\bibitem[{\citenamefont{Muthukumar}(1997)}]{Dynamics1997Muthukumar}
\bibinfo{author}{\bibfnamefont{M.}~\bibnamefont{Muthukumar}},
  \bibinfo{journal}{J. Chem. Phys.} \textbf{\bibinfo{volume}{107}},
  \bibinfo{pages}{2619} (\bibinfo{year}{1997}).

\bibitem[{\citenamefont{Shin et~al.}(2009)\citenamefont{Shin, Drew, Bartles,
  Wong, and Grason}}]{Cooperativity2009Shin}
\bibinfo{author}{\bibfnamefont{H.}~\bibnamefont{Shin}},
  \bibinfo{author}{\bibfnamefont{K.~R.~P.} \bibnamefont{Drew}},
  \bibinfo{author}{\bibfnamefont{J.~R.} \bibnamefont{Bartles}},
  \bibinfo{author}{\bibfnamefont{G.~C.~L.} \bibnamefont{Wong}},
  \bibnamefont{and} \bibinfo{author}{\bibfnamefont{G.~M.}
  \bibnamefont{Grason}}, \bibinfo{journal}{Phys. Rev. Lett.}
  \textbf{\bibinfo{volume}{103}}, \bibinfo{pages}{238102}
  (\bibinfo{year}{2009}).

\bibitem[{\citenamefont{Shin and Grason}(2010)}]{Structural2010Shin}
\bibinfo{author}{\bibfnamefont{H.}~\bibnamefont{Shin}} \bibnamefont{and}
  \bibinfo{author}{\bibfnamefont{G.~M.} \bibnamefont{Grason}},
  \bibinfo{journal}{Phys. Rev. E} \textbf{\bibinfo{volume}{82}},
  \bibinfo{pages}{051919+} (\bibinfo{year}{2010}).

\bibitem[{\citenamefont{Bloomfield}(1996)}]{DNA1996Bloomfield}
\bibinfo{author}{\bibfnamefont{V.~A.} \bibnamefont{Bloomfield}},
  \bibinfo{journal}{Curr. Opin. Struct. Biol.} \textbf{\bibinfo{volume}{6}},
  \bibinfo{pages}{334} (\bibinfo{year}{1996}).

\bibitem[{\citenamefont{Kornyshev and Leikin}(1998)}]{Symmetry1998Kornyshev}
\bibinfo{author}{\bibfnamefont{A.~A.} \bibnamefont{Kornyshev}}
  \bibnamefont{and} \bibinfo{author}{\bibfnamefont{S.}~\bibnamefont{Leikin}},
  \bibinfo{journal}{Biophys. J.} \textbf{\bibinfo{volume}{75}},
  \bibinfo{pages}{2513} (\bibinfo{year}{1998}).

\bibitem[{\citenamefont{Kornyshev and Leikin}(1997)}]{Theory1997Kornyshev}
\bibinfo{author}{\bibfnamefont{A.~A.} \bibnamefont{Kornyshev}}
  \bibnamefont{and} \bibinfo{author}{\bibfnamefont{S.}~\bibnamefont{Leikin}},
  \bibinfo{journal}{J. Chem. Phys.} \textbf{\bibinfo{volume}{107}},
  \bibinfo{pages}{3656} (\bibinfo{year}{1997}).

\bibitem[{\citenamefont{Holmes et~al.}(1990)\citenamefont{Holmes, Popp,
  Gebhard, and Kabsch}}]{Atomic1990Holm}
\bibinfo{author}{\bibfnamefont{K.~C.} \bibnamefont{Holmes}},
  \bibinfo{author}{\bibfnamefont{D.}~\bibnamefont{Popp}},
  \bibinfo{author}{\bibfnamefont{W.}~\bibnamefont{Gebhard}}, \bibnamefont{and}
  \bibinfo{author}{\bibfnamefont{W.}~\bibnamefont{Kabsch}},
  \bibinfo{journal}{Nature} \textbf{\bibinfo{volume}{347}}, \bibinfo{pages}{44}
  (\bibinfo{year}{1990}).

\bibitem[{\citenamefont{McLaughlin et~al.}(1993)\citenamefont{McLaughlin,
  Gooch, Mannherz, and Weeds}}]{Structure1993McLaughlin}
\bibinfo{author}{\bibfnamefont{P.~J.} \bibnamefont{McLaughlin}},
  \bibinfo{author}{\bibfnamefont{J.~T.} \bibnamefont{Gooch}},
  \bibinfo{author}{\bibfnamefont{H.~G.} \bibnamefont{Mannherz}},
  \bibnamefont{and} \bibinfo{author}{\bibfnamefont{A.~G.} \bibnamefont{Weeds}},
  \bibinfo{journal}{Nature} \textbf{\bibinfo{volume}{364}},
  \bibinfo{pages}{685} (\bibinfo{year}{1993}).

\bibitem[{\citenamefont{Al-Khayat et~al.}(1995)\citenamefont{Al-Khayat, Yagi,
  and Squire}}]{Structural1995AlKhayat}
\bibinfo{author}{\bibfnamefont{H.~A.} \bibnamefont{Al-Khayat}},
  \bibinfo{author}{\bibfnamefont{N.}~\bibnamefont{Yagi}}, \bibnamefont{and}
  \bibinfo{author}{\bibfnamefont{J.~M.} \bibnamefont{Squire}},
  \bibinfo{journal}{J. Mol. Biol.} \textbf{\bibinfo{volume}{252}},
  \bibinfo{pages}{611} (\bibinfo{year}{1995}).

\bibitem[{\citenamefont{Tsuda et~al.}(1996)\citenamefont{Tsuda, Yasutake,
  Ishijima, and Yanagida}}]{Torsional1996Tsuda}
\bibinfo{author}{\bibfnamefont{Y.}~\bibnamefont{Tsuda}},
  \bibinfo{author}{\bibfnamefont{H.}~\bibnamefont{Yasutake}},
  \bibinfo{author}{\bibfnamefont{A.}~\bibnamefont{Ishijima}}, \bibnamefont{and}
  \bibinfo{author}{\bibfnamefont{T.}~\bibnamefont{Yanagida}},
  \bibinfo{journal}{Proc. Natl. Acad. Sci. USA} \textbf{\bibinfo{volume}{93}},
  \bibinfo{pages}{12937} (\bibinfo{year}{1996}).

\bibitem[{\citenamefont{Noé et~al.}(2008)\citenamefont{Noé, Daidone, Smith,
  di~Nola, and Amadei}}]{Solvent2008Noe}
\bibinfo{author}{\bibfnamefont{F.}~\bibnamefont{Noé}},
  \bibinfo{author}{\bibfnamefont{I.}~\bibnamefont{Daidone}},
  \bibinfo{author}{\bibfnamefont{J.~C.} \bibnamefont{Smith}},
  \bibinfo{author}{\bibfnamefont{A.}~\bibnamefont{di~Nola}}, \bibnamefont{and}
  \bibinfo{author}{\bibfnamefont{A.}~\bibnamefont{Amadei}},
  \bibinfo{journal}{J. Phys. Chem. B} \textbf{\bibinfo{volume}{112}},
  \bibinfo{pages}{11155} (\bibinfo{year}{2008}).

\bibitem[{\citenamefont{Hribar et~al.}(2002)\citenamefont{Hribar, Southall,
  Vlachy, and Dill}}]{How2002Hribar}
\bibinfo{author}{\bibfnamefont{B.}~\bibnamefont{Hribar}},
  \bibinfo{author}{\bibfnamefont{N.~T.} \bibnamefont{Southall}},
  \bibinfo{author}{\bibfnamefont{V.}~\bibnamefont{Vlachy}}, \bibnamefont{and}
  \bibinfo{author}{\bibfnamefont{K.~A.} \bibnamefont{Dill}},
  \bibinfo{journal}{J. Am. Chem. Soc.} \textbf{\bibinfo{volume}{124}},
  \bibinfo{pages}{12302} (\bibinfo{year}{2002}).

\bibitem[{\citenamefont{Tulip and Bates}(2009)}]{Peptide2009Tulip}
\bibinfo{author}{\bibfnamefont{P.~R.} \bibnamefont{Tulip}} \bibnamefont{and}
  \bibinfo{author}{\bibfnamefont{S.~P.} \bibnamefont{Bates}},
  \bibinfo{journal}{The Journal of Chemical Physics}
  \textbf{\bibinfo{volume}{131}}, \bibinfo{pages}{015103+}
  (\bibinfo{year}{2009}).

\bibitem[{\citenamefont{Limbach et~al.}(2006)\citenamefont{Limbach, Arnold,
  Mann, and Holm}}]{Espresso2006Limbach}
\bibinfo{author}{\bibfnamefont{H.}~\bibnamefont{Limbach}},
  \bibinfo{author}{\bibfnamefont{A.}~\bibnamefont{Arnold}},
  \bibinfo{author}{\bibfnamefont{B.}~\bibnamefont{Mann}}, \bibnamefont{and}
  \bibinfo{author}{\bibfnamefont{C.}~\bibnamefont{Holm}},
  \bibinfo{journal}{Comput. Phys. Commun.} \textbf{\bibinfo{volume}{174}},
  \bibinfo{pages}{704} (\bibinfo{year}{2006}).

\bibitem[{\citenamefont{Egelman et~al.}(1982)\citenamefont{Egelman, Francis,
  and DeRosier}}]{Factin1982Egelman}
\bibinfo{author}{\bibfnamefont{E.~H.} \bibnamefont{Egelman}},
  \bibinfo{author}{\bibfnamefont{N.}~\bibnamefont{Francis}}, \bibnamefont{and}
  \bibinfo{author}{\bibfnamefont{D.~J.} \bibnamefont{DeRosier}},
  \bibinfo{journal}{Nature} \textbf{\bibinfo{volume}{298}},
  \bibinfo{pages}{131} (\bibinfo{year}{1982}).

\bibitem[{\citenamefont{Splettstoesser
  et~al.}(2011)\citenamefont{Splettstoesser, Holmes, No\'{e}, and
  Smith}}]{Structural2011Splettstoesser}
\bibinfo{author}{\bibfnamefont{T.}~\bibnamefont{Splettstoesser}},
  \bibinfo{author}{\bibfnamefont{K.~C.} \bibnamefont{Holmes}},
  \bibinfo{author}{\bibfnamefont{F.}~\bibnamefont{No\'{e}}}, \bibnamefont{and}
  \bibinfo{author}{\bibfnamefont{J.~C.} \bibnamefont{Smith}},
  \bibinfo{journal}{Proteins: Structure, Function, and Bioinformatics}
  \textbf{\bibinfo{volume}{79}}, \bibinfo{pages}{2033} (\bibinfo{year}{2011}).

\bibitem[{\citenamefont{Chu and Voth}(2006)}]{Coarse2006Chu}
\bibinfo{author}{\bibfnamefont{J.-W.} \bibnamefont{Chu}} \bibnamefont{and}
  \bibinfo{author}{\bibfnamefont{G.~A.} \bibnamefont{Voth}},
  \bibinfo{journal}{Biophysical journal} \textbf{\bibinfo{volume}{90}},
  \bibinfo{pages}{1572} (\bibinfo{year}{2006}).

\bibitem[{\citenamefont{Jho et~al.}(2008)\citenamefont{Jho, Kandu\v{c}, Naji,
  Podgornik, Kim, and Pincus}}]{Strong2008Jho}
\bibinfo{author}{\bibfnamefont{Y.~S.} \bibnamefont{Jho}},
  \bibinfo{author}{\bibfnamefont{M.}~\bibnamefont{Kandu\v{c}}},
  \bibinfo{author}{\bibfnamefont{A.}~\bibnamefont{Naji}},
  \bibinfo{author}{\bibfnamefont{R.}~\bibnamefont{Podgornik}},
  \bibinfo{author}{\bibfnamefont{M.~W.} \bibnamefont{Kim}}, \bibnamefont{and}
  \bibinfo{author}{\bibfnamefont{P.~A.} \bibnamefont{Pincus}},
  \bibinfo{journal}{Phys. Rev. Lett.} \textbf{\bibinfo{volume}{101}},
  \bibinfo{pages}{188101} (\bibinfo{year}{2008}).

\end{thebibliography}
\bibliographystyle{rsc} 
}

\end{document}